\documentclass[%
reprint,
amsmath,amssymb,
aps,
]{revtex4-2}

\usepackage{graphicx}
\usepackage{dcolumn}
\usepackage{bm}
\usepackage[per-mode = reciprocal]{siunitx}
\usepackage{amsmath}
\usepackage{cases}
\usepackage{xcolor}

\newcommand{\nocontentsline}[3]{}
\newcommand{\tocless}[2]{\bgroup\let\addcontentsline=\nocontentsline#1{#2}\egroup}

\begin{document}


\title{Controlled manipulation of solitons in a recirculating fiber loop using external potentials}

\author{François Copie}
\email{francois.copie@univ-lille.fr}
\author{Pierre Suret}%
\author{Stéphane Randoux}%

\affiliation{%
	Univ. Lille, CNRS, UMR 8523 - PhLAM - Physique des Lasers Atomes et Molécules, F-59000 Lille, France
}%

\begin{abstract} 
Optical solitons are self-sustained wave packets that propagate without distortion due to a balance between dispersion and nonlinearity. Their unique stability underpins key photonic applications while also playing a central role in nonlinear wave physics. However, real-time control over soliton dynamics in non-dissipative systems remains a major challenge, limiting their practical applications in photonic systems. Here, we introduce a fiber-based platform for soliton manipulation, by creating programmable external potentials through synchronous arbitrary phase modulation in a recirculating optical fiber loop. We demonstrate precise soliton trapping, parametric excitation, and coupled multi-soliton interactions, revealing particle-like behavior in excellent agreement with a Hamiltonian description in which solitons are treated as interacting particles. The strong analogy with matter-wave solitons in Bose-Einstein condensates highlights the broader implications of our approach, which provides a versatile experimental tool for the study of nonlinear wave dynamics and engineered soliton manipulation.

\end{abstract}

\maketitle

\tocless\section{Introduction}
Solitons are localised coherent nonlinear structures involved in the complex dynamical behaviour of many nonlinear wave systems \cite{yang_nonlinear_2010}. Their ubiquitous nature is reflected in the fact that they occur in very diverse fields including hydrodynamics, plasmas, quantum gas physics or nonlinear optics. Their robustness underscores their significance in practical systems such as mode-locked lasers, supercontinuum sources \cite{dudley_supercontinuum_2010}, and frequency combs, and they have also been extensively explored for high-speed data transmission \cite{hasegawa_solitons_1995}.

In their simplest -- conservative -- form, solitons are persisting wavepackets that retain their shape and velocity over virtually infinite time owing to the balance between dispersive and nonlinear effects. From a mathematical perspective, these solitons represent exact solutions of the class of integrable partial differential equations which includes the 1D nonlinear Schrödinger (NLS) equation. This equation that describes, at leading order, the propagation of light fields in optical fibers can be solved using the inverse scattering transform (IST) method \cite{yang_nonlinear_2010}. Within this framework, solitons are identified as nonlinear eigenmodes that can be used in practical applications such as fiber telecommunications \cite{turitsyn_nonlinear_2017}. A key feature of their dynamics is the ability to interact or \textit{collide} elastically, exhibiting particle-like behaviour \cite{yang_nonlinear_2010}. Eventhough the exact dynamics of the optical field during the collision process is complex, the only effect of the interaction that is visible afterwards is a positional shift of the colliding solitons \cite{zakharov_interaction_1973, yang_nonlinear_2010} which has often been interpreted as resulting from an interaction force \cite{gordon_interaction_1983, mitschke_experimental_1987}. In an integrable setting, however, soliton interactions are highly constrained, and external control over their motion is impossible. This implies that the spatiotemporal dynamics of a set of interacting solitons, no matter how complex, is completely determined by the initial wavefield and further manipulation during evolution is essentially prohibited. The lack of a general method to dynamically manipulate solitons' trajectories during propagation limits their potential for tunable optical applications as well as more fundamental investigations of complex many soliton interaction.

In contrast, matter-wave solitons in Bose-Einstein condensates (BECs) can be formed and subsequently manipulated using external optical potentials \cite{strecker_formation_2002, khaykovich_formation_2002}. In this context, the degree of freedom introduced by the application of an external potential has proven particularly promising in manipulating the spatiotemporal dynamics of solitons, enabling, for instance, their repeated collision \cite{nguyen_collisions_2014} or interaction with barriers that can lead to their controlled splitting \cite{marchant_controlled_2013, wales_splitting_2020}. Technical constraints in generating stable arbitrary optical traps and physical constraints such as spatial dimensionality or the collapse instability might however hinder part of this potential. Achieving an equivalent level of control with optical solitons would significantly benefit the study of nonlinear wave physics and could unlock novel functionalities for soliton-based applications.

In this work, we present a versatile fiber-optic system for soliton manipulation, based on synchronous phase modulation within a recirculating loop. By applying a configurable spatiotemporal external potential to propagating solitons, we demonstrate soliton trapping, parametric excitation, and controlled multi-soliton interactions confirming the particle-like nature of solitons in our experiments. Our approach establishes a novel experimental platform for exploring and engineering nonlinear optical wave dynamics while further strengthening the existing parallel with matter-wave systems.

Our system relies on the recirculating optical fiber loop configuration which gives access to the space-time dynamics of the wavefield \cite{desurvire_raman_1985, mollenauer_experimental_1990, nakazawa_10_1991, goossens_experimental_2019} enabling single-shot observation of non-repetitive dynamics. This has recently been used to observe the nonlinear spatiotemporal dynamics of modulation instability \cite{kraych_nonlinear_2019, kraych_statistical_2019, copie_spatiotemporal_2022}, soliton collision \cite{copie_spacetime_2023} and soliton gases \cite{suret_soliton_2023, fache_perturbed_2025}. This approach stands in contrast to most optical methods for accessing space-time dynamics, which typically rely on indirect techniques -- such as parameter scanning to simulate different propagation distances \cite{pierangeli_observation_2018, marcucci_topological_2019, xin_evidence_2021, dieli_observation_2024, van_simaeys_experimental_2001}, backscattered signal processing \cite{mussot_fibre_2018}, or iterative reconstructions \cite{sheveleva_idealized_2022} -- and are generally limited to deterministic dynamics.\\[1em]

\tocless\section{\label{sec:setup} Experimental setup and models}

\begin{figure}[]
	\center \includegraphics[width=.48\textwidth]{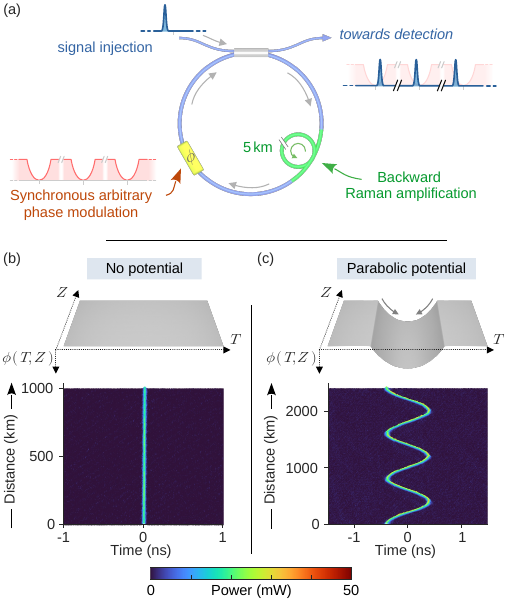}
	\caption{\label{fig:setup} \textbf{Experimental setup}. (a) Principle of operation of the synchronously phase modulated recirculating fiber loop. Example of experimentally recorded spatiotemporal dynamics of a single soliton in the case of (b) no external potential and (c) trapping within a truncated parabolic potential (used for describing the setup in (a)). Typical peak power and duration of the solitons are $\sim \SI{30}{mW}$ and $\sim \SI{45}{ps}$ FWHM.}
\end{figure}

The recirculating loop platform implemented in this work is entirely fiber based and its principle of operation is schematically depicted in Fig.\,\ref{fig:setup}(a). A single frequency, continuous wave (cw) laser at \SI{1555}{nm} is modulated in intensity through a \SI{20}{GHz} bandwidth electro-optic modulator to generate the initial optical signal which consists of a custom pattern of pulses that realizes a specific set of experiments. The electric signal that creates the pulses pattern is obtained using the 1\textsuperscript{st} channel of a \SI{13.5}{GHz} bandwidth arbitrary waveform generator (AWG). An acousto-optic modulator is then used to ensure a high extinction of surrounding waves and to prevent the buildup of Brillouin scattering. This field is fed to the recirculating loop through a 90/10 coupler whose ports are arranged such that 90\% of the power is recirculating. Consequently, 10\% of the circulating field is extracted at each roundtrip and directed towards the detection system.

The recirculating loop itself measures approximately \SI{5}{km} (roundtrip time $\approx \SI{24.552}{\micro\s}$) and is mostly comprised of a spool of commercial SMF-28 fiber [group velocity dispersion (GVD) coefficient $\beta_2 = \SI{-22}{ps^2/km}$, Kerr nonlinearity coefficient $\gamma = \SI{1.23}{ \per\W \per\km}$] connected at both ends to wavelength division multiplexers (WDMs). Through these, light from a \SI{1455}{nm} pump laser is coupled in and out of the loop to realize backward Raman amplification. The laser's power is gated to the duration of the experiment and finely tuned to compensate for the total dissipation experienced by the signal during a roundtrip. In this way, losses are effectively canceled for the signal which can then propagate over extremely long distances, typically several thousands of kilometers. A representative evolution of the optical power during a single experiment is given in Supplement. We emphasize that in our experiments, even though the optical power is on average almost constant, the amplitude of the optical wavefield may undergo significant evolution in the course of one roundtrip due to the distributed Raman amplification. Upon consecutive circulations, the wavefield thus experiences periodic evolution of its power. For the range of parameters used in this work the typical linear and nonlinear length scales associated to the pulses we consider are $L_\text{L} \approx L_\text{NL} \approx \SI{30}{km}$ which is substantially longer than the length of a roundtrip (see parameters above and in the caption of Fig.\,\ref{fig:setup}). In this case it is known that the nonlinear pulses behave as \textit{effective} solitons also referred to as \textit{guiding-center} solitons \cite{hasegawa_guiding-center_1990, hasegawa_guiding-center_1991}.     

We create an effective external potential by using an electro-optic phase modulator $\text{EOM}_\phi$ embedded in the fiber loop which is driven synchronously with both the intensity pulse pattern and the roundtrip time of the loop. This ensures precise and repeatable initial positioning of the pulses relative to the potential and prevents drifting of the latter relative to the circulating field during propagation. Prior to the experiment, a mapping of the targeted potential landscape is designed which consists of a sequence of waveforms (one per roundtrip). The 2\textsuperscript{nd} channel of the AWG sequentially plays each waveform to generate the electrical signal that drives $\text{EOM}_\phi$ at each pass of the optical signal. In the general case, a different waveform can be played at each roundtrip, resulting in a dynamic potential landscape. The AWG can also be set to play the same waveform and thus create a static potential. Note that intra-loop phase modulation has already been used to realize experiments in synthetic frequency dimensions \cite{dutt_experimental_2019, chen_real-time_2021, englebert_bloch_2023, cheng_multi-dimensional_2023, senanian_programmable_2023} and generate optical gap solitons \cite{bersch_optical_2012}, typically relying on static harmonic modulations. For the latter, the potential consists of a fast modulation which creates an array of coupled waveguides shorter than the wavepacket itself. By contrast, our experiments operate in a different parameter regime in which the potential timescale is broader than the soliton duration. Application of an external potential is also a powerful means of tailoring the properties of Kerr cavity solitons in resonators \cite{englebert_manipulation_2024, sun_dissipative_2022}. In that context, however, phase modulation is constrained by the requirement for the waveform to match specific cavity resonances, typically limiting it to periodic shapes. These constraints do not apply in our non-resonant loop system, allowing arbitrary and dynamically varying phase profiles. In a different context, it is noteworthy that static periodic potentials have also been proposed as supporting structures to stabilize the propagation of soliton trains for long-haul transmissions, although in that case the potential was generated via cross-phase modulation with an optical clock signal \cite{widdowson_soliton_1994, bigo_error-free_1997, shipulin_suppression_1997}.

The system uses essentially polarization maintaining (PM) fiber components except for the \SI{5}{km} fiber spool and the WDMs. A polarization controller is thus placed after this non-PM section to realign the polarization of the signal after each circulation which mitigates polarization dependent effects.   

The signal extracted from the loop at the 10\% port of the coupler is detected with a \SI{65}{GHz} bandwidth oscilloscope used in a sequential mode: after each roundtrip, the oscilloscope is triggered to record the signal in a segment. The full recording consists of up to a few thousands segments and thus contains the dynamics of the signal for more than \SI{10000}{km} of propagation. As an illustration, two typical examples of fully processed recordings are shown in Fig.\,\ref{fig:setup}(b, c) depicting respectively the spatiotemporal dynamics of a single soliton freely propagating and oscillating within a parabolic potential generated via synchronous phase modulation with a truncated quadratic profile. Additional information concerning the experimental setup and details regarding optical power calibration are given in Supplement.
\vspace{1em}

The evolution of the wavefield in the recirculating fiber loop over several roundtrips is well captured by the following 1D NLS equation also referred to as 1D Gross-Pitaevskii (GP) equation in the Bose-Einstein condensate community

\begin{equation}\label{eq:NLSE_mf}
	i\frac{\partial A}{\partial Z} = \frac{\beta_2}{2} \frac{\partial^2 A}{\partial T^2} - \gamma |A|^2A - i\frac{\alpha_\text{eff}}{2}A - \frac{\phi(T, Z)}{L}A,
\end{equation}

where $A(T, Z)$ is the envelope of the electromagnetic field, $Z$ is the propagation distance, $T$ is the time defined in the reference frame traveling at the group velocity at the carrier frequency. $\beta_2$ and $\gamma$ are the group velocity dispersion (GVD) and the Kerr nonlinearity coefficients respectively. $\alpha_\text{eff}$ is a linear dissipation term that describes the potentially imperfect loss compensation which results in an effective exponential power variation over many roundtrips. $L$ is the total length of the recirculating loop. $\phi(T, Z)$ is the phase modulation applied at distance $Z = nL$ where $n$ is the roundtrip number. As explicitly shown in Supplement, Eq. (\ref{eq:NLSE_mf}) derives from an iterative model accounting for the nonlinear light propagation in the loop and the periodic application of discrete recirculating conditions (local losses and phase modulation). It is valid provided that the roundtrip-to-roundtrip evolution of the field envelope $A_n$ and the phase modulation $\phi_n$ are small. Consequently, the subscript $n$ can be dropped, and the term $\phi(T, Z)/L$ plays the role of an external potential that results from the longitudinal distribution of the phase modulation $\phi(T, Z)$. It is noteworthy in this model that, since $\phi$ can now be seen as a continuous function of $T$ and $Z$, the phase modulation applied in a discrete fashion (once per roundtrip) generates a distributed potential landscape. We checked numerically that in the range of parameters of all the experiments reported in this work the mean-field and iterative models yield identical results. The signal measured in our experiments is proportional to the optical power sampled at each roundtrip which corresponds to $|A(T, Z = nL)|^2$.

It is possible to write Eq.\,(\ref{eq:NLSE_mf}) in a normalized form which, considering the anomalous dispersion regime (i.e. $\beta_2 < 0$) reads 
\begin{equation}\label{eq:NLS_adim}
	i\dfrac{\partial \psi}{\partial t} = -\frac{1}{2}\dfrac{\partial^2\psi}{\partial x^2} - |\psi|^2\psi - i\epsilon\psi + V(x, t) \psi, 
\end{equation}

\noindent by introducing the following adimensional variables: $\psi = A/\sqrt{P_0}$; $x = T\sqrt{\gamma P_0/|\beta_2|}$; $t = Z \gamma P_0$; $\epsilon = \alpha_\text{eff}/(2\gamma P_0)$; $V = \textcolor{red}{-}\phi/(\gamma L P_0)$. Note that within this notation, the physical propagation distance $Z$ (resp. physical time $T$) is replaced by a normalised time $t$ (resp. normalised position $x$).
\vspace{1em}

\begin{figure}[]
	\center \includegraphics[width=.47\textwidth]{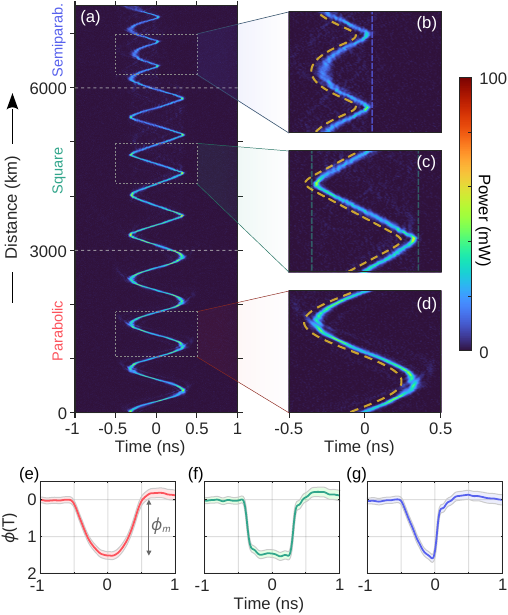}
	\caption{\label{fig:static} \textbf{Dynamics of a single soliton in a trapping potential taking consecutively three different shapes}. (a) Space-time dynamics of a single soliton subjected to the potential trap. Zooms highlighting (d) sinusoidal oscillations, (c) total temporal reflections and (b) succession of temporal reflections and parabolic trajectories. Dashed lines in (b-d) are the trajectories estimated by the particle model shifted by $\SI{-80}{ps}$ for clarity. (e-g) Shapes of the phase modulations applied in each segment obtained from heterodyne measurement (see Supplement for details regarding the measurement method). Solid lines are the averaged profiles and colored areas show the standard deviation of the phase measurement.}
\end{figure}

Considering that individual NLS solitons behave like particles, maintaining their shape at all positions in space and time, a simplified Hamiltonian description has been proposed in Refs.\,\cite{scharf_soliton_1992, martin_bright_2007, martin_bright_2008}. In this model, the interaction among solitons is described by an intersoliton potential that has been introduced in order to match the shift experienced by solitons upon pairwise collisions. Considering an initial wavefield consisting of $N$ well-separated single solitons, the initial condition reads: 
\begin{equation}\label{eq:SolCI}
	\psi(x, t = 0) = \sum_{i = 1}^{N} 2\eta_i \text{sech}[2\eta_i(x - q_{0i})] e^{i(v_{0i} x + \phi_{0i})}.
\end{equation}
\noindent Each soliton is characterized by a set of 4 parameters: $\eta_i$, $q_{0i}$, $v_{0i}$, and $\phi_{0i}$, encoding its amplitude, initial position, velocity and phase respectively. In the particle analogy, the Hamiltonian $H$ for such system reads \cite{scharf_soliton_1992, martin_bright_2007, martin_bright_2008}
\begin{multline}\label{eq:H}
	H = \sum_{i = 1}^{N} \frac{p_i^2}{2\eta_i} + \eta_i V(q_i) \\ - \sum_{1 \leq i < j \leq N} 2\eta_i \eta_j (\eta_i + \eta_j) \text{sech}^2\left[ \frac{2\eta_i \eta_j}{\eta_i + \eta_j} (q_i - q_j) \right] ,
\end{multline}
\noindent where $q_i(t)$ denotes the position of the peak of soliton $i$ and $p_i(t) = \eta_i \dot{q_i}$ its momentum. The first two terms represent the kinetic and potential energies, while the last term corresponds to the soliton-soliton interaction potential. This expression is valid for arbitrary potential shapes provided that $V(q)$ varies slowly compared to the soliton width, ensuring that each soliton experiences the potential through its local value $V(q_i)$. Note that the phase of the solitons does not appear in Eq.\,(\ref{eq:H}) since it does not affect the positional shift induced by the collisions. The relative phase between solitons however influences the exact spatiotemporal dynamics observed during collisions \cite{copie_spacetime_2023}. The simultaneous interaction of more than two solitons is not accounted for by this Hamiltonian, limiting its accuracy only to scenarios involving pairwise collisions occurring at distinct positions in space and time. Note also that this model fails to reproduce the correct shift when solitons interact with vanishing relative velocities \cite{scharf_soliton_1992}. In this work, we have solved the Hamilton’s equation of motion $\dot{q_i} = \partial H / \partial p_i$ and $\dot{p_i} = -\partial H / \partial q_i$ to find the trajectories of solitons $q_i(t)$, within the particle analogy, subjected to external potentials $V(x, t)$ and compared them to our experimental observations.\\[1em]

\tocless\section{\label{sec:exp} Results}

\tocless\subsection{\label{subsec:ssdyn} Single soliton dynamics}

\begin{figure}[]
	\center \includegraphics[width=.47\textwidth]{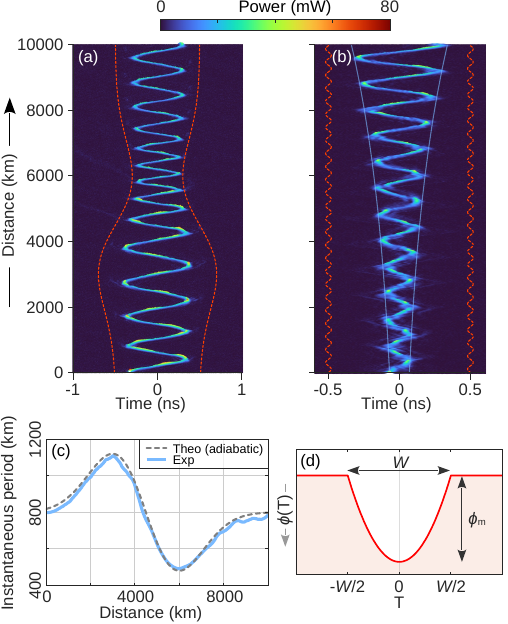}
	\caption{\label{fig:dynamic} \textbf{Manipulation of solitons using spatiotemporally varying harmonic potentials.} (a) Soliton in a slowly modulated trap and (b) driven at the parametric resonance. The red dashed lines denote the boundaries of the truncated parabolic traps. Outside of the region enclosed by these lines the external potential is flat. The blue lines in (b) indicate the exponential growth of the amplitude of the oscillations predicted by the particle model. (c) Evolution of the instantaneous period of oscillation of the soliton in (a) compared to the prediction of Eq.\,(\ref{eq:period}) considering an adiabatic evolution. (d) Sketch of the truncated parabolic phase modulation.}
\end{figure}

The first set of experiments that we present focuses on the spatiotemporal dynamics of a single soliton under a trapping potential that sequentially takes three distinct shapes: a quadratic well, a square box, and an asymmetric well. In the spatiotemporal diagram of Fig.\,\ref{fig:static}(a) the horizontal axis is the time in the reference frame that moves at the group velocity of the signal in absence of external potential. The external potential is set to be stationary in this reference frame. The zero of this axis is the central position of the external potential and the soliton is initially shifted by \SI{-300}{ps} from the center [see Fig.\,\ref{fig:static}(a) and \ref{fig:static}(e)]. In the first \SI{3000}{km} of propagation a truncated parabolic phase modulation spanning \SI{1}{ns} is synchronously applied which creates an effective quadratic potential. As a result, the soliton's velocity is modulated and its temporal position oscillates sinusoidally [Fig.\,\ref{fig:static}(d)] similar to what is observed in BEC experiments \cite{nguyen_collisions_2014, wales_splitting_2020}. It should be noted that this oscillation of the velocity physically translates into an oscillation of the central frequency of the Fourier spectrum of the soliton. Little distortion of the soliton's shape is observed as a consequence of the interaction with weak radiative waves that are also trapped but this was not found to alter its trajectory in a significant way. Considering a quadratic phase modulation of the form $\phi(T)/L = -C/2 \times T^2$, the trap wavenumber is determined, by analogy with the quantum harmonic oscillator, to be $K_0^2 = \beta_2 C$. When applying a truncated quadratic phase modulation parametrized by its depth in radian $\phi_m$ and duration (width) $W$, as schematically shown in Fig.\,\ref{fig:dynamic}(d), the spatial period of oscillation $Z_0$ can be expressed 

\begin{equation}\label{eq:period}
	Z_0 = \frac{2\pi}{K_0} = \sqrt{\frac{\pi^2W^2L}{2 |\beta_2| \phi_m}}.
\end{equation}

The experimentally observed period $Z_0 = \SI{835}{km}$ is consistent with the phase modulation depth obtained via heterodyne measurement $\phi_m \approx 1.5 \pm \SI{0.1}{rad}$ [Fig.\,\ref{fig:static}(e)].
After \SI{3000}{km} of propagation, the phase modulation is abruptly switched (from one roundtrip to the next) to produce a square-well potential. As a result, the spatiotemporal dynamics of the soliton is modified: it experiences total reflections at the edges while crossing the flat bottom of the potential at a constant velocity without changing shape [Fig.\,\ref{fig:static}(c)]. After each reflection, the soliton recovers precisely its shape. The observed dynamics is characteristic of the  temporal reflection phenomenon which is known to be the analog of reflection of an optical beam at the interface between two dielectric media \cite{xiao_reflection_2014, plansinis_what_2015}.
At a propagation distance of \SI{6000}{km}, the phase modulation is switched to generate an asymmetric, semiparabolic well in which the soliton experiences temporal reflection at the sharp right boundary and follows a curved trajectory in the region of quadratic potential [Fig.\,\ref{fig:static}(b)].

The results illustrated in Fig.\,\ref{fig:static} demonstrate the capacity of our experimental setup to manipulate individual 1D solitons through the application of an arbitrary shaped and reconfigurable external potential. The trajectory of the soliton in response to each potential shape is perfectly reproduced within the particle analogy by integrating the simplified Hamiltonian model [Eq.\,(\ref{eq:H})] for a single particle subjected to corresponding potentials. This is illustrated in Fig.\,\ref{fig:static}(b--d) where the dashed lines (shifted for clarity) represent the trajectories numerically computed and rescaled to physical units.

The recirculating loop configuration allows the phase modulation signal applied to the circulating field to be updated at each roundtrip. This enables the creation of dynamic potential landscapes (external potentials varying in both $T$ and $Z$), providing an additional tool for controlling optical waves. As a demonstration, we consider a parabolic trap with $Z$-dependent strength, with results presented in Fig.\,\ref{fig:dynamic}. In Fig.\,\ref{fig:dynamic}(a) and (b), the parabolic trap strength is longitudinally modulated by varying its temporal width $W$ while keeping its depth $\phi_m$ constant [schematized in Fig.\,\ref{fig:dynamic}(d)]. In both cases, the evolution of $W$ follows a continuous function sampled once per roundtrip. However, the variation of $W$ from roundtrip to roundtrip is small, ensuring that the wavefield experiences a smoothly evolving potential landscape without discernible effects of the discrete nature of the phase modulation. When the trap is slowly modulated [Fig.\,\ref{fig:dynamic}(a)], the soliton's oscillation period and amplitude increase as $W$ increases, and vice versa. Under adiabatic evolution, Eq.\,(\ref{eq:period}) indeed consistently predicts that the soliton oscillation period is proportional to $W$. The experimentally measured evolution of the period aligns well with this prediction for adiabatic evolution [Fig.\,\ref{fig:dynamic}(c)], showcasing precise control over the spatiotemporal dynamics of trapped solitons. This concept extends to scenarios involving multiple interacting solitons. Figure \ref{fig:dynamic}(b) illustrates another example of soliton control, achieved by periodically modulating the spatial frequency of the harmonic potential at parametric resonance. It has been shown theoretically that the motion of a soliton under such conditions follows the equation of motion of the parametrically driven pendulum \cite{kartashov_parametric_2004, tenorio_dynamics_2005}: a modulation period close to half the oscillator's natural period induces exponential growth in oscillation amplitude. This is realized in our case using a truncated parabolic trap weakly modulated according to $W(Z) = W_0(1 - 0.05\sin(4\pi Z /Z_0))$ with $W_0 = \SI{1}{ns}$ and  $Z_0 = \SI{800}{km}$. The oscillations of the soliton initially near the trap's center (\SI{-75}{ps}) exhibit rapid amplitude growth due to parametric excitation. Our experimental results strongly agree with the particle model, as shown by the amplitude of the oscillations closely following the predicted exponential amplification [blue lines in Fig.\,\ref{fig:dynamic}(b)]. It should be noted that we focus here on the positional dynamics of the pulse, but the presence of the trapping potential eventually introduces internal dynamics in the form of width oscillations, which may lead to more complex behaviour \cite{baizakov_double_2005}.

The superposition of the predicted trajectories over the experimental observations reported in Fig.\,\ref{fig:dynamic}(a, b) is provided in the Supplement, showing an excellent agreement. 

\tocless\subsection{\label{subsec:inter_soliton} Interacting solitons}

\begin{figure*}[]
	\center \includegraphics[width=.97\textwidth]{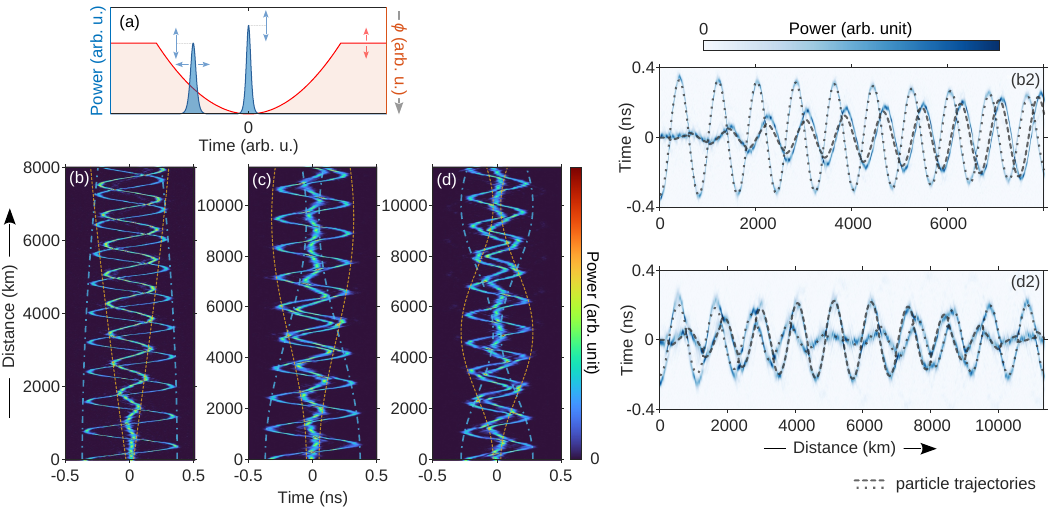}
	\caption{\label{fig:2sol} \textbf{Coherent dynamics of soliton pairs.} (a) Schematic view of the initial condition of the experiments: one soliton is initially centered in the quadratic trap while the other is detuned. (b-d) Spatiotemporal dynamics recorded for three increasing coupling strengths: $\Delta Z/ \ Z_0 = -1.6\%$, $-6.9\%$, $-11.2\%$ respectively. Dashed blue and orange lines denote the amplitude of oscillation of each soliton. (b2) and (d2) Same as (b) and (d) rotated by 90°. Simulated particle trajectories obtained from Eq.\,(\ref{eq:H}) are superimposed as dashed and dotted lines.}           
\end{figure*}

The experiments presented thus far all involve a single soliton subjected to an effective external potential whose primary effect is to modulate the velocity of the soliton. We now consider the scenario in which several solitons are confined in a stationary quadratic trapping potential which makes it possible to produce intricate multi-collision dynamics. Such scenario is usually precluded in nonlinear fiber optics experiments as solitons with different velocities can only interact once in absence of potential. We focus on a regime in which the solitons are well separated and distinguishable before and after interaction (i.e. sufficiently broad trapping potential and large initial separation between solitons). Under these conditions, the solitons dynamics can be modeled to a good approximation using the Hamiltonian particle model [Eq.\,(\ref{eq:H})].

We first present experiments realized in the two-solitons configuration for which the particle model is integrable which has been shown to imply regular (non-chaotic) dynamics of the system \cite{martin_bright_2007}. To better emphasize the dynamics of the soliton pair, we designed initial conditions consisting of a soliton placed at the center of the harmonic trap while a second one is generated shifted from the center [situation depicted in Fig.\,\ref{fig:2sol}(a)]. In absence of interaction, the centered soliton would remain pinned to the center of the trap. In contrast, the coherent interaction with the second soliton results in the soliton pair exhibiting coupled dynamics. We varied the strength of the coupling by changing the natural period of the trap (phase modulation depth) and the characteristics of the solitons (amplitude and initial temporal separation) as illustrated schematically in Fig.\,\ref{fig:2sol}(a).

We show in Fig.\,\ref{fig:2sol}(b--d) three experiments sorted from weakest to strongest coupling. Over the range of propagation distances plotted, the two solitons collide in each realization at least 20 times without significant alteration of their initial amplitude, duration and shape. The dashed colored lines serve as visual guides for tracking the dynamics of each soliton. For the weakest coupling [Fig.\,\ref{fig:2sol}(b)], the central soliton is gradually set into motion and the amplitude of the oscillations of its trajectory increases at the expense of that of the initially shifted soliton. Further increasing the coupling strength [Fig.\,\ref{fig:2sol}(c)], we observe after $\sim\SI{9000}{km}$ a point at which the amplitude of oscillation of the shifted soliton is canceled whereas the initially centered one oscillates strongly. Figure \ref{fig:2sol}(d) shows the most strongly coupled dynamics where the solitons apparently exchange their initial position twice, reaching back almost perfectly the initial configuration after \SI{10000}{km}. Although the trajectory of each soliton is less distinguishable because of their smaller separation, their coupled motion is still remarkably regular.

A convenient way to quantify the coupling strength which has been reported in Ref.\,\cite{nguyen_collisions_2014} is to measure the relative shift of the period of oscillation of the solitons with respect to the natural period $Z_0$ of the trap (i.e. the period of oscillation experienced by a single soliton within the same trap) $\Delta Z_0/ \ Z_0$. The collision process shifts solitons forward in their propagation direction such that the coupling between solitons effectively decreases their period of oscillation \cite{martin_collision-induced_2016}. Using reference single soliton experiments performed simultaneously to those reported in Fig.\,\ref{fig:2sol}(b--d) but not shown for clarity, we were able to estimate these shifts as $\Delta Z_0/ \ Z_0 = -1.6\%$, $-6.9\%$, $-11.2\%$ respectively. The latter is roughly 4 times larger than the matter-wave soliton experiments reported in Ref.\,\cite{nguyen_collisions_2014}.

Our experiments can be compared to the predictions of the Hamiltonian model which treats solitons as interacting classical particles. By integrating Hamiltonian dynamical equations with the Hamiltonian given by Eq.\,(\ref{eq:H}), we obtain the positional dynamics $q_1(t)$ and $q_2(t)$ of the two equivalent particles associated with the solitons of amplitude $\eta_1$ and $\eta_2$ respectively. The computed trajectories, rescaled to physical units, corresponding to the cases illustrated in Fig.\,\ref{fig:2sol}(b) and (d) are superimposed on the experiments in Fig.\,\ref{fig:2sol}(b2) and (d2) and  reproduce quantitatively very well the coupled dynamics of the soliton pairs. For better visualization, the spatiotemporal representations are rotated by 90° and the colormap has been changed. It should be noted that the initial temporal detuning between the solitons in the particle model was lowered slightly relative to the experiments to account for a transient initial regime during which the external potential stabilizes. 

\begin{figure}[]
	\center \includegraphics[width=.48\textwidth]{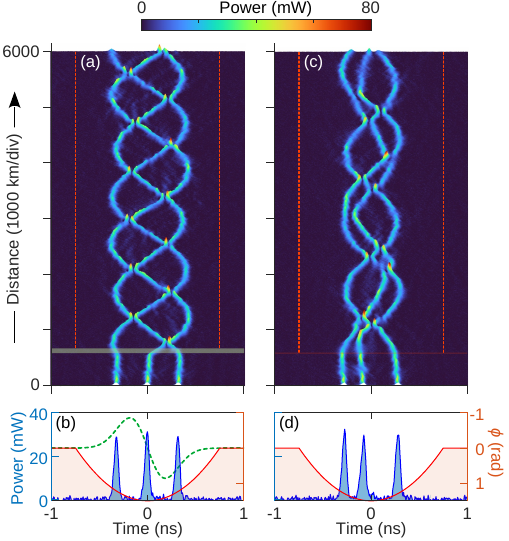}
	\caption{\label{fig:35sol} \textbf{Regular and chaotic dynamics of soliton triplets.} Spatiotemporal dynamics of a (a) regular and (c) chaotic soliton triplet. The green horizontal line in (a) shows the range of distance over which velocities of the solitons are initiated. Dashed vertical lines indicate the boundaries of the truncated parabolic trap. (b, d) Respective intensity profiles recorded at $z = \SI{500}{km}$ in blue. Solid red lines and shaded areas illustrate the trapping potential applied after \SI{500}{km} while the dashed green line in (b) indicates the phase modulation applied for $\SI{500}{km} \leq z \leq \SI{585}{km}$ to initiate velocities of the solitons.}
\end{figure}

Finally, we realized the same type of experiment with three solitons interacting in a harmonic trap. In practice, the number of solitons and their initial temporal position is easily set by tailoring the electric signal sent to the intensity EOM. The three-particle system is not integrable and supports both chaotic and regular regimes \cite{martin_bright_2007, martin_bright_2008}. We demonstrate here [see Fig.\,\ref{fig:35sol}(a)] that we are able to specifically generate soliton triplets with a regular dynamics by carefully shaping the initial state, i.e. the position and relative velocity of each soliton. To do so, three equally spaced solitons are first generated and propagate freely for \SI{500}{km} before a specifically shaped phase modulation is applied for a few consecutive roundtrips. The profile of the phase modulation [dashed green line in Fig.\,\ref{fig:35sol}(b)] is specifically designed such that the central soliton acquires a positive velocity while the two surrounding ones get a velocity of opposite sign and half amplitude. This in turn produces a triplet of solitons characterised by regular pairwise collisons when the trapping harmonic potential is turned on.

Figure \ref{fig:35sol}(c) shows an example of chaotic motion produced following a similar procedure. The central soliton is here slightly detuned and no relative velocity is imprinted before the trap is turned on. As a result, the coupled dynamics becomes very intricate, with in particular the three solitons overlapping at their first encounter. In this configuration especially, the particle model is not able to describe properly the observed dynamics and exhibits an increased sensitivity to the initial setting (see Supplement for a related discussion). These experiments provide a clear demonstration of the capabilities offered by this experimental system in designing arbitrary multi-soliton optical wavepackets. In particular, the results presented in Fig.\,\ref{fig:35sol} shows how our system can be a powerful experimental platform for exploration of chaotic dynamics in wave mechanics. \\[1em]

\tocless\section{\label{sec:conclusion} Conclusion}
In summary, we have experimentally realized the spatiotemporal manipulation of NLS solitons within a recirculating optical fiber loop. The implementation of intra-loop synchronous phase modulation enables the generation of an effective external potential acting on the trajectories of solitons. We take advantage of the capacity to arbitrary shape the potential landscape in both space and time to demonstrate various control scenarios including periodic oscillation, temporal reflection and parametric excitation. In all cases, the solitons showcase particle-like behaviors with trajectories akin to that of classical particles. In the situation where two solitons collide repeatedly within a trapping potential, the collective motion of the system is also accurately described by a simplified Hamiltonian model treating solitons as interacting particles. Precise design of the initial wavefield and external potential reveals the predicted onset of chaotic motion when more than two solitons are interacting.

The photonic platform presented here benefits from a high degree of configurability in addition to truly single-shot operation, which makes it readily compatible for more advanced manipulation of optical wavefields. In particular, it constitutes a promising approach towards shaping of the inverse scattering transform (IST) ``nonlinear'' spectrum of complex multi-soliton waveforms with possible implications for nonlinear ``eigenvalue'' telecommunications \cite{turitsyn_nonlinear_2017, mucci_manipulation_2025}.

The weak breaking of integrability provided by an external potential offers a possible route towards the observation of thermalization of soliton gases, i.e. large random ensembles of solitons \cite{suret_soliton_2024}. In this sense, our system proves highly valuable for testing and refining the predictions of the recent theoretical framework of generalized hydrodynamics (hydrodynamics of many-body integrable systems), which, to date,  has been almost exclusively focused on quantum systems (\cite{doyon_generalized_2025} and references therein, \cite{bastianello_observation_2025}). Localization and branched flow phenomena are also expected to occur within periodic \cite{scharf_length-scale_1993} or random \cite{patsyk_observation_2020, patsyk_incoherent_2022} potential landscapes respectively. The study of the scattering of nonlinear waves from potential barriers is one fundamental topic that could be addressed, and which can greatly benefit from the strong analogy between nonlinear optics and matter-waves. Although there is a wealth of literature on this topic, practical implementations in matter-wave contexts are limited \cite{ernst_resonant_2010, helm_splitting_2014, marchant_quantum_2016, wales_splitting_2020, lorenzi_atomic_2024}. Many of these may be more easily realized using photonic platforms.
\vspace{1em}

\noindent \textbf{Acknowledgment.} The authors are grateful to H. Damart, G. Dekyndt, and the Centre d’Etudes et de Recherche Lasers et Application (CERLA) for technical support. This work has been partially supported by the Agence Nationale de la Recherche through the StormWave (ANR-21-CE30-0009) and SOGOOD (ANR-21-CE30-0061) projects, the LABEX CEMPI project (ANR-11-LABX-0007), the Ministry of Higher Education and Research, Hauts de France council and European Regional Development Fund (ERDF) through the Contrat de Projets Etat-Région (CPER Photonics for Society P4S).\vspace{.5em}

\noindent \textbf{Data Availability.} Data underlying the results presented in this paper are available in Ref.\,\cite{copie_data_2025}. \vspace{.5em}





%


\clearpage

\renewcommand{\theequation}{S\arabic{equation}}
\setcounter{figure}{0}
\setcounter{equation}{0}
\onecolumngrid

\renewcommand{\theequation}{S\arabic{equation}}
\renewcommand{\thefigure}{S\arabic{figure}}

\begin{center}
	{\bf Supplemental material for : \\"Controlled manipulation of solitons in a recirculating fiber loop using external potentials"}\\
\end{center}

\begin{center}
	François Copie, Pierre Suret, and St\'ephane Randoux
\end{center}

\begin{center}
	{\it Univ. Lille, CNRS, UMR 8523 - PhLAM -Physique des Lasers Atomes et Molécules, F-59000 Lille, France}
\end{center}

\noindent 
	In this Supplemental Material we provide some details about the experimental setup, the experimental methodology and the mathematical models describing our system. It also provides additional comparison between experiments and the Hamiltonian model describing solitons as classical particles. We also discuss annex question about the failure of the particle model in the case where the interaction among 3 solitons occur at the same space-time position. All equations, figures, reference numbers within this document are prepended with “S” to distinguish them from corresponding numbers in the main manuscript.

\setcounter{tocdepth}{1}
\tableofcontents
\vspace{2em}

\setcounter{section}{0}

\section{Details of the experimental system}

\subsection{Description of the experimental setup}

The recirculating loop platform implemented in this work is entirely fiber based and is schematically depicted in Fig.\,\ref{fig:setup}(a). The operating principle consists in three stages that are described here: (i) \textit{Generation of the initial optical wavefield}; (ii) \textit{Propagation in the recirculating loop and application of the external potential}; (iii) \textit{Single-shot detection}.\vspace{1em}

\noindent{(i) \textbf{\textit{Generation of the initial optical wavefield}}}\vspace{.5em}

\noindent  A single frequency, continuous wave (cw) laser at \SI{1555}{nm} is modulated in intensity using a \SI{20}{GHz} bandwidth electro-optic modulator (EOM\textsubscript{\textit{I}} Exail MXER-LN-20) to generate a \SI{200}{ns} long optical signal which consists of a custom pattern of pulses that realizes a specific set of experiments (illustrated in the top right corner of Fig.\,\ref{fig:setup}(a)). The electric signal that creates the pulses pattern is obtained using a dual-channel \SI{13.5}{GHz} bandwidth arbitrary waveform generator (Tektronix AWG70002B). A single run contains a few tens of completely programmable independent experiments involving short solitonic pulses along with a sequence of pulses solely used for power calibration and synchronization purposes. This pattern is continuously generated and amplified using an Erbium doped fiber amplifier (EDFA) to reach Watt-level peak power. An acousto-optic modulator (AOM) is used to pick a pulses pattern that will constitute the initial wavefield for the experiment and ensures a high extinction of surrounding waves to prevent the buildup of Brillouin scattering. The AOM is driven by the signal from an arbitrary waveform generator (AFG\textsubscript{1}). The field is then fed to the recirculating loop through a 90/10 coupler whose ports are arranged such that 90\% of the power is recirculating. Consequently, 10\% of the circulating field is extracted at each roundtrip and directed towards the detection system.\vspace{1em}

\noindent{(ii) \textbf{\textit{Propagation in the recirculating loop and application of the external potential}}}\vspace{.5em}

\noindent The recirculating loop itself measures approximately \SI{5}{km} (roundtrip time $\approx \SI{24.552}{\micro\s}$) and is mostly comprised of a spool of commercial SMF-28 fiber (GVD coefficient $\beta_2 = \SI{-22}{ps^2/km}$, Kerr nonlinearity coefficient $\gamma = \SI{1.23}{ \per\W \per\km}$) connected at both ends to wavelength division multiplexers (WDMs). Through these, light from a \SI{1455}{nm} pump laser (Keopsys CRFL) is coupled in and out of the loop to realize backward Raman amplification. The laser's power is gated to the duration of the experiment using an AOM and finely tuned with a mechanical variable attenuator (VA) to compensate for the total dissipation experienced by the signal during a roundtrip. This way, losses are effectively canceled for the signal which can then propagate over extremely long distances. Figure \ref{fig:setup}(b) shows the normalised optical power evolution in a single experiment, namely, the regular three-solitons dynamics illustrated in Fig.\,5(a) of the main manuscript.

\begin{figure}[!h]
	\includegraphics[width=.9\textwidth]{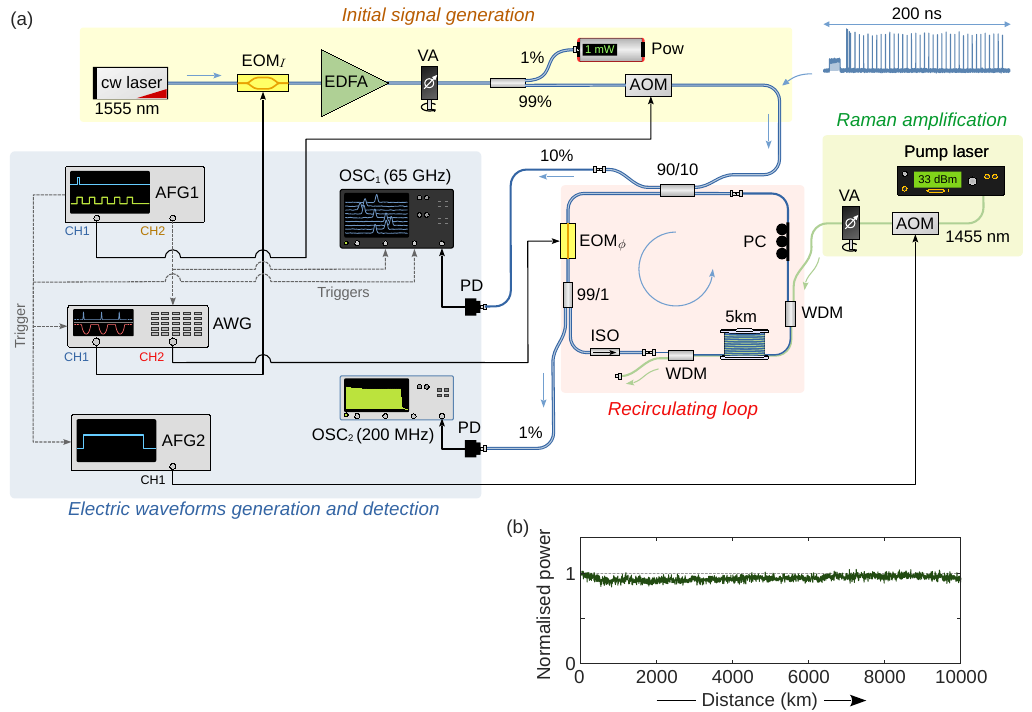}
	\caption{\label{fig:setup} \textbf{Experimental system}. (a) Schematic of the setup. (b) Longitudinal normalised evolution of the optical power in a single experiment. AFG: arbitrary function generator, AOM: acousto-optic modulator, AWG; arbitrary waveform generator, cw: continuous wave, EDFA: Erbium doped fiber amplifier, EOM: electro-optic modulator, ISO: isolator, OSC: oscilloscope, PC: polarisation controller, PD: photodector, Pow: power meter, VA: variable attenuator, WDM: wavelength division multiplexer.}
\end{figure}

An effective external potential is obtained by using an electro-optic phase modulator (EOM\textsubscript{$\phi$} Exail MPZ-LN-20) embedded in the fiber loop. Prior to the experiment, a mapping of the targeted potential landscape is designed which consists of a sequence of waveforms (one per roundtrip). The 2\textsuperscript{nd} channel of the AWG sequentially plays each waveform to generate the electrical signal that drives $\text{EOM}_\phi$ at each pass of the signal. To do so, the AWG is triggered by a burst signal from AFG\textsubscript{1} whose period is finely tuned to match the roundtrip time of the fiber loop. In the general case, a different waveform can be played at each roundtrip resulting in a dynamic potential landscape. The AWG can also be set to play the same waveform and thus create a static potential.

A tap coupler extracts 1\% of the field which is used to monitor in real-time the evolution of the optical power to finely tune the parameters of the Raman amplification. Note that the system uses essentially polarization maintaining (PM) fiber components except for the \SI{5}{km} fiber spool and the WDMs. A polarization controller is thus placed after this non-PM section to realign the polarization of the signal after each circulation which mitigates polarization dependent effects.\vspace{1em}

\noindent{(iii) \textbf{\textit{Single-shot detection}}}\vspace{.5em}

The signal extracted from the loop at the 10\% port of the main coupler is detected using a \SI{50}{GHz} bandwidth photodiode (Finisar XPDV2120R) coupled to a \SI{65}{GHz} bandwidth oscilloscope (LeCroy Labmaster 10-Zi-A) used in a sequential mode: after each roundtrip, the oscilloscope is triggered to record the pattern of pulses in a segment. The full recording consists of up to a few thousands segments and can thus contains the dynamics of the signal for more than \SI{10000}{km} of propagation. Importantly, the same burst signal that triggers application of the synchronous phase modulation is used to trigger the acquisition of each segment, such that the optical signal is naturally recorded in the reference frame of the external potential.

\subsection{Voltage-to-Optical Power conversion}

The signal recorded consists of $\sim 10\%$ of the optical power extracted from the recirculating fiber loop at each roundtrip. The optical signal is detected and converted to an electric voltage via an XPDV2120R \SI{50}{GHz} photodetector coupled to a Lecroy Labmaster 10 Zi-A \SI{65}{GHz} oscilloscope. A calibration is thus needed to convert the measured voltage into optical power circulating within the fiber loop. This consists in the determination of a conversion factor that is applied a posteriori during the processing of the experimental data. The calibration procedure takes advantage of the process of spontaneous modulation instability (MI) that occurs in the propagation of flat-top pulses. This fundamental nonlinear phenomenon materializes in the destabilization of an initially noisy quasi-continuous pulse which entails the formation at intermediate propagation distance of a quasi-periodic pulse train. The small signal gain associated to the process reads

\begin{equation} \label{eq:gain}
	g(\Delta f) = 4\pi^2 |\beta_2 \Delta f|\sqrt{\Delta f_\text{c}^2 - \Delta f^2}
\end{equation}

\noindent with $\Delta f_\text{c} = \frac{1}{\pi}\sqrt{\gamma P_0/\beta_2}$. $\Delta f$ is the relative frequency detuning with respect to the pump frequency, $\beta_2$ and $\gamma$ are the group velocity dispersion coefficient and Kerr nonlinearity respectively, and $P_0$ is the average optical power of the initial quasi-continuous pump. The Fourier spectrum of the optical waves sees the growth of sidebands symmetrically located around the pump according to the gain curve (\ref{eq:gain}) with power-dependent cutoffs at $\pm \Delta f_\text{c}$ and peak positions $\pm \Delta f_\text{MI} = \pm \Delta f_\text{c}/\sqrt{2}$.

\begin{figure}[h]
	\centering
	\includegraphics[width=.85\linewidth]{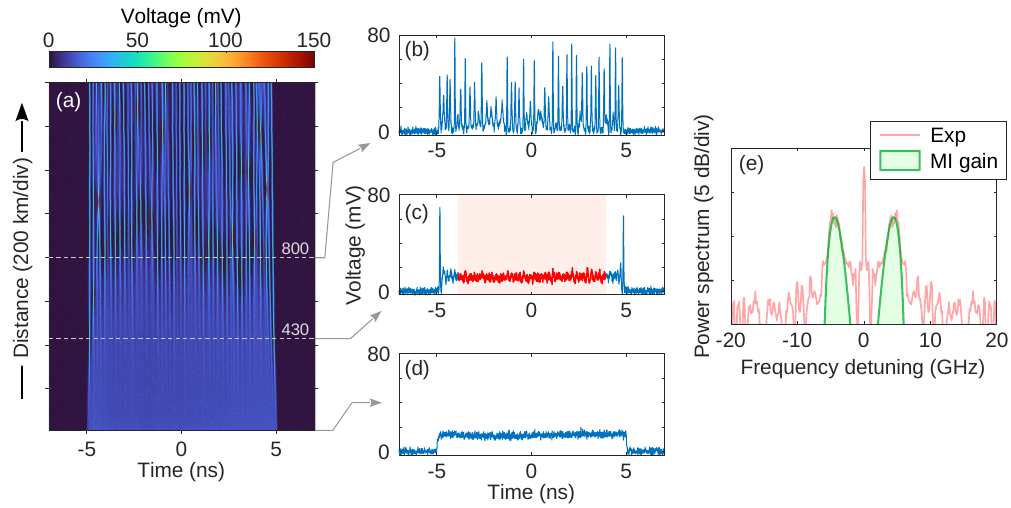}
	\caption{\textbf{Calibration procedure for Volt-to-Optical Power conversion based on spontaneous MI}. (a) Experimentally recorded spatiotemporal diagram of spontaneous MI on a \SI{10}{ns}-long flat-top pulse. (b-d) Voltage recorded at $z = \SI{800}{km}$, $\SI{430}{km}$, and $\SI{0}{km}$ showing the fully developed MI pattern, earlier stage spontaneous modulation, and initial pulse profile respectively. (e) FFT analysis of the flat-top part of the pulse at $z = \SI{430}{km}$ in red and fitted small signal MI gain spectrum in green.}
	\label{fig:calib}
\end{figure}

The initial signals used in our experiments consists in arbitrarily designed optical patterns enabling several independent experimental runs to be realized in the same recording. Alongside the experimental runs, a portion of the pattern is dedicated to the power calibration and the jitter correction procedure. Taking for instance the recording associated to Fig.\,4 of the main manuscript, a flat-top pulse of \SI{10}{ns}-long duration is generated and after \SI{800}{km} of propagation, it has turned into a high contrast train of pulses as a consequence of spontaneous MI (see Fig.\,\ref{fig:calib}(a, b)). The Fourier spectrum of the optical signal at an earlier stage ($z_\text{cal} = \SI{430}{km}$, Fig.\,\ref{fig:calib}(c)), exhibits MI induced sidebands with peak position $|\Delta f_\text{MI}| \approx \SI{4.5}{GHz}$. The small signal MI gain spectrum $\exp[g(\Delta f)z_\text{cal}]$ can be adjusted to the experimental spectrum with $P_0$ as fitting parameter. Figure \ref{fig:calib}(e) shows the best fit for this recording obtained for $P_0 = \SI{7}{mW}$. The mean voltage recorded at the first roundtrip for the flat-top part of the pulse is $V_0 = \SI{13.5}{mW}$ (Fig.\,\ref{fig:calib}(d)). The Volt-to-Optical Power conversion factor is consequently $P_0/V_0 = \SI{0.52}{W/V}$ for this specific recording. The voltage measured experimentally is then simply multiplied by this factor to be converted to optical power. 

This procedure is performed systematically for each recording and shows negligible variation.
\vspace{2em}

\subsection{Measurement of the phase modulation profiles}

The electro-optic phase modulator (EOM\textsubscript{$\phi$}) embedded within the fiber loop enables application of arbitrary shaped phase modulation synchronous to the circulating field resulting in an effective external potential. We measured the profile of the phase modulations independently of the soliton propagation experiments reported in this work by performing heterodyne measurements described here. This is useful for two main reasons: (i) Confirm the arbitrary shape of the external potentials (i.e. minimal distortion with respect to the targeted shape) ; (ii) Separately confirm the depth of the potential traps which can otherwise be inferred directly in the experiments from the oscillation period of the solitons.

\begin{figure}[h]
	\centering
	\includegraphics[width=.75\linewidth]{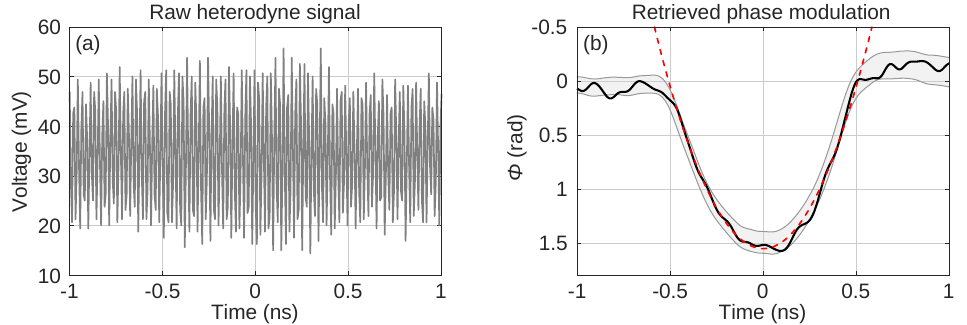}
	\caption{\textbf{Heterodyne measurement of the phase modulation}. (a) Sample of raw heterodyne recording in the case of quadratic modulation and (b) corresponding single-shot phase modulation retrieved. The gray shaded area illustrates the standard deviation computed from 1000 single-shot recordings. The dashed red line is an illustrative quadratic fit. The average profile in plotted in Fig.\,2(e) of the manuscript.}
	\label{fig:phase}
\end{figure}

An initial condition consisting in a long optical pulse in generated and sent to the recirculating fiber loop. During the first circulation, the phase modulation to be characterized is applied to the flat-top part of the pulse. Heterodyne detection is realized by mixing the optical signal exiting the fiber loop with a continuous wave laser detuned by $\Delta f = \SI{-45}{GHz}$ (accordingly $\Delta \lambda \sim\SI{+0.36}{nm}$) before being recorded by the same detection system as the one used for the experiments. The applied phase modulation is retrieved from the recording via a simple numerical procedure: (i) frequency components centered around the beating frequency are selected using a passband filter and centered around null frequency; (ii) in the time domain, the phase is obtained as the argument of the complex array containing the filtered data. This single-shot phase measurement is repeated 1000 times which enables computation of a mean profile and standard deviation as shown in Fig.\,2(e-g) of the manuscript. Figure \ref{fig:phase}(a, b) shows an example of raw heterodyne signal in the case of quadratic modulation and the corresponding single-shot phase profile retrieved respectively.

\newpage

\section{Derivation of the 1D-Gross-Pitaevskii mean-field equation from the iterative map}

\subsection*{Iterative model in physical units} \label{sec:Phys}

Experiments realised in the recirculating fiber loop are described by an iterative model consisting of a propagation equation and a boundary (recirculating) condition.

\begin{numcases}{}
	\label{eq:NLSE_it}
	\dfrac{\partial A_n}{\partial Z} = -i\left( \dfrac{\beta_2}{2} \dfrac{\partial^2 A_n}{\partial T^2} - \gamma |A_n|^2 A_n \right) + \dfrac{g_\text{r}}{2} A_n\\[1em]
	\label{eq:rec_cond}
	A_{n+1}(T, Z = 0) = \rho A_n(T, Z = L)~\text{e}^{i\phi_n(T)}
\end{numcases}

Propagation over the roundtrip $n$ follows a NLSE with gain [Eq.\,(\ref{eq:NLSE_it})]. The field at the beginning of roundtrip $n + 1$ as a function of the one at the end of roundtrip $n$ is given by Eq.\,(\ref{eq:rec_cond}). $\rho$ is a constant coefficient that account for local losses.

\begin{equation}\label{eq:lumped_losses}
	\rho = \text{e}^{-\frac{(g_\text{r} + \alpha_\text{eff})}{2}L}
\end{equation}

In this expression, $g_\text{r}$ account for the distributed Raman amplification while $\alpha_\text{eff}$ traduces the imperfect loss compensation which results in an effective exponential power decay. $L$ is the length of the recirculating loop.

The phase modulation $\phi_n(T)$ can additionally be applied and appears in the recirculating condition.

\begin{figure}[!h]
	\centering
	\includegraphics[width=0.55\linewidth]{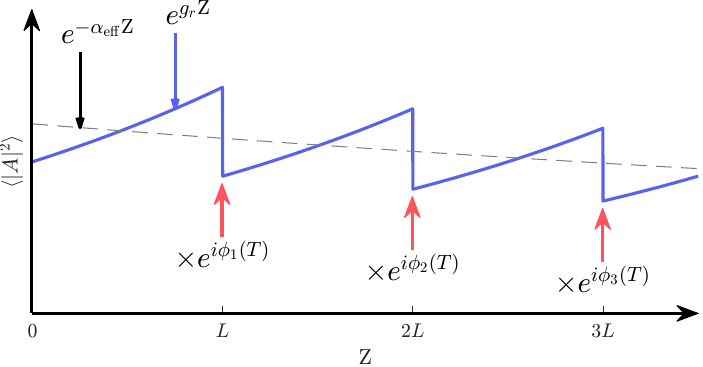}
	\caption{\textbf{Illustration of the power evolution in the iterative model}. The blue curve illustrates the periodic amplification and local losses encountered across consecutive roundtrips while the gray dashed line illustrates a potential slow effective evolution of the mean power. Periodic application of a phase modulation at each roundtrip is depicted by the red arrows.}
	\label{fig:it_model}
\end{figure}

Figure \ref{fig:it_model} schematically depicts the power evolution over 3 roundtrips.

\subsection*{Lumped model}

Equations (\ref{eq:NLSE_it}) and (\ref{eq:rec_cond}) can be reduced to a single PDE by expressing the discrete recirculating condition as a function of a Dirac delta comb.\\

\begin{equation}\label{eq:diff_form_sum}
	\frac{\partial A_n}{\partial Z}\bigg|_{Z=nL} = \sum_n \ln \left(\rho ~\text{e}^{i\phi_n(T)} \right) \delta(Z - nL) A_n
\end{equation}

Including the boundary condition (\ref{eq:rec_cond}) into (\ref{eq:NLSE_it}) consists in adding the right hand side of (\ref{eq:diff_form_sum}) to (\ref{eq:NLSE_it}):

\begin{equation}\label{eq:single_PDE}
	\dfrac{\partial A_n}{\partial Z} = -i\left( \dfrac{\beta_2}{2} \dfrac{\partial^2 A_n}{\partial T^2} - \gamma |A_n|^2 A_n \right) + \dfrac{g_\text{r}}{2} A_n + \sum_n \left(-\frac{g_r + \alpha_\text{eff}}{2}L + i\phi_n(T) \right) \delta(Z - nL) A_n,
\end{equation}

\noindent where the expression of $\rho$ (\ref{eq:lumped_losses}) has been explicited.

\subsection*{Mean-field model}

Integrating Eq.\,(\ref{eq:single_PDE}) from $z_{n-} = z_n - \epsilon$ to $z_{n-} = z_n + \epsilon$, where $0 < \epsilon \ll 1$ and assuming continuity of $A$ and its temporal derivatives we find that the field $A_n(z_{n+})$ just after the boundary is expressed as a function of the one just before $A_n(z_{n-})$ by 

\begin{equation}\label{eq:bc2}
	A_n(z_{n+}) = A_n(z_{n-}) \left[1 -\frac{g_r + \alpha_\text{eff}}{2}L + i\phi_n(T) \right] ,
\end{equation}

which is a rewriting of Eq.\,(\ref{eq:rec_cond}).\\

Assuming small variation of the field $A$ during propagation from $Z = 0$ to $Z = L$, i.e. in the course of a roundtrip, we can write (first order Taylor expansion):

\begin{equation}\label{eq:bc3}
	A_{n+1}(0, T) = \underbrace{\left( A_n(0, T) + \frac{\partial A_n}{\partial Z}\bigg|_{Z = 0} \times L \right)}_{A_n(L, T)} \left[1 -\frac{g_r + \alpha_\text{eff}}{2}L + i\phi_n(T) \right],
\end{equation}

\noindent by noting that $A_n(z_{n-})$ corresponds to $A_n(L, T)$ and $A_n(z_{n+})$ corresponds to $A_{n+1}(0, T)$.

The partial derivative which describes the small variation of $A$ during one roundtrip is then evaluated with Eq.\,(\ref{eq:NLSE_it}). By keeping only first order terms and identifying $\frac{\partial A}{\partial Z} = (A_{n+1} - A_{n})/L$ we find the mean-field model

\begin{equation}\label{eq:mfmodel}
	\boxed{
		\frac{\partial A}{\partial Z} = -i\dfrac{\beta_2}{2} \dfrac{\partial^2 A}{\partial T^2} +  i\gamma |A|^2 A - \frac{\alpha_\text{eff}}{2}A + i\frac{\phi(Z, T)}{L}A,
	}
\end{equation}

\vspace{1em} \noindent which is Eq.\,(1) of the manuscript. This takes the form of a damped NLSE with an external potential played by the discrete phase modulation that is distributed. The roundtrip number $n$ has been dropped and $Z$ is the propagation distance that is not restricted to the roundtrip length $[0; L]$ anymore. It should be noted that the model remains valid for phase modulation varying slowly from roundtrip to roundtrip which enables a $Z$ dependence of $\phi$ provided that $\partial\phi / \partial Z \ll 1$.   

\section{Comparison between experiments and Hamiltonian model in dynamic potential}

In the manuscript we present experimental recordings of the dynamics of single solitons subjected to dynamically varying external potentials, namely a truncated parabolic trap with $Z$-dependent strength. In these experiments, the depth of the phase modulation is fixed at $\sim \SI{1.75}{rad}$ and the strength is modulated by varying the duration $W$. Here, we compare the observed trajectories of the solitons reported in this article to the corresponding predictions of the Hamiltonian model.

\begin{figure}[h]
	\centering
	\includegraphics[width=.85\linewidth]{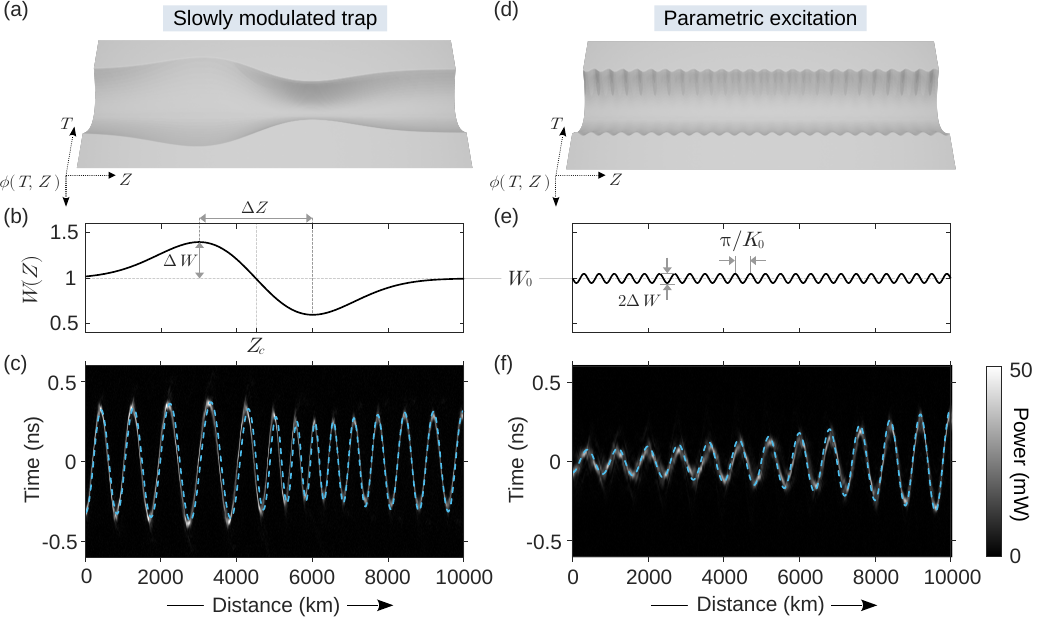}
	\caption{\textbf{Dynamically modulated potentials: Experiments vs. Hamiltonian model}. (a-c) Slowly modulated and (d-f) parametrically modulated parabolic traps. (a, d) 3D views of the external potentials, (b, e) Longitudinal evolutions of the parabolic trap duration $W(Z)$. (c, f) Spatiotemporal dynamics recorded experimentally and particle trajectories predicted by the Hamiltonian model superimposed as dashed blue lines.}
	\label{fig:dyn_trap}
\end{figure}

In the case of the slowly modulated trap illustrated in Fig.\,\ref{fig:dyn_trap}(c), the duration of the truncated modulation $W$ follows the expression:

\begin{equation}
	W(Z) = W_0 - 2\frac{\Delta W}{\Delta Z} \exp(0.5) \times (Z - Z_c)\exp\left[\frac{\sqrt{2}(Z - Z_c)}{\Delta Z}\right]^2,
	\label{eq:WZ}
\end{equation}

\noindent with $W_0 = \SI{1}{ns}$, $\Delta W = \SI{0.4}{ns}$, $\Delta Z = \SI{3000}{km}$ and $Z_c = \SI{4500}{km}$. This evolution is depicted in Fig.\,\ref{fig:dyn_trap}(b) and Fig.\,3(a) of the manuscript (dashed red lines). From Eq.\,(\ref{eq:WZ}) and Eq.\,(5) of the main manuscript it is possible to compute the evolution of the trap wavenumber $K_0(Z)$ which is then used to compute the particle trajectory in the Hamiltonian model. The result is superimposed on the experiment as a dashed blue line in Fig.\,\ref{fig:dyn_trap}(c) and shows excellent agreement.

In the case of parametrically modulated trap illustrated in Fig.\,\ref{fig:dyn_trap}(f), the duration of the truncated modulation $W$ follows the expression:

\begin{equation}
	W(Z) = W_0\left[1 - \frac{\Delta W}{W_0}\sin\left(2 K_0 Z \right) \right],
	\label{eq:pr}
\end{equation}

\noindent with $W_0 = \SI{1}{ns}$, $\Delta W = \SI{50}{ps}$ and $K_0 = 2\pi / Z_0$ with the natural spatial period of the trap $Z_0 = \SI{800}{km}$. The factor 2 in the sine function denotes the parametric forcing at half the natural period of the trap. This evolution is depicted in Fig.\,\ref{fig:dyn_trap}(e) and Fig.\,3(b) of the manuscript (dashed red lines). Again, Eq.\,(\ref{eq:pr}) and Eq.\,(5) of the main manuscript are used to compute the evolution of the trap wavenumber $K_0(Z)$ then used to compute the particle trajectory in the Hamiltonian model. The result is superimposed on the experiment as a dashed blue line in Fig.\,\ref{fig:dyn_trap}(f) and shows a remarkable agreement, in particular in reproducing the exponential growth of the oscillation amplitude.\\

\section{Sensitivity of the Hamiltonian model with three trapped particles}

In the manuscript we report the observation of regular and chaotic three-solitons dynamics in a harmonic trap (see Fig.\,5) and mention the failure of the Hamiltonian model in the latter case because of the simultaneous interaction of the 3 ``particles'', a situation mentioned in Ref.\,\cite{scharf_soliton_1992}. Additionally, the discrepancy between the model and an actual realization (physical or numerical) builds up from the accumulation of small error originating from the fact that trapped solitons do not reach asymptotic separation after collision as discussed in Refs.\,\cite{martin_bright_2007, martin_bright_2008}. This is all the more significant when larger number of solitons are in strong interaction.

Here we illustrate how a quantitative agreement between our observations and the particle model is particularly hard to reach in the three-solitons situation by comparing the experimental result with predicted particle trajectories with 3 slightly different initial conditions.

\begin{figure}[h]
	\centering
	\includegraphics[width=.85\linewidth]{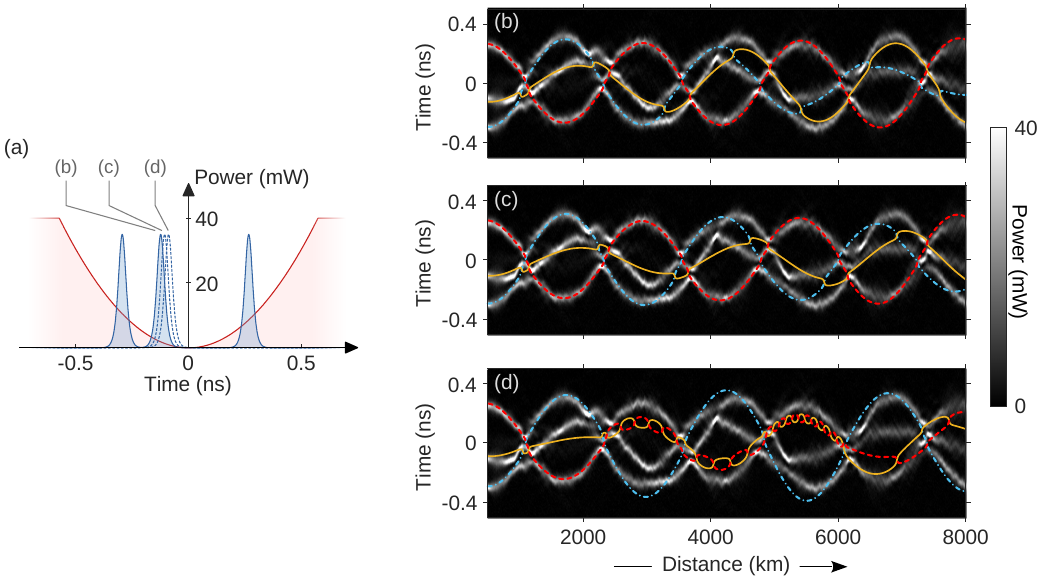}
	\caption{\textbf{Three soliton dynamics: Experiment vs. Hamiltonian model}. (a) Sketch of the configuration when the trapping potential is turned on ($z = \SI{500}{km}$). (b-d) Spatiotemporal dynamics recorded experimentally with the 3 computed particle trajectories superimposed as colored lines. The 3 panels show trajectories computed for 3 slightly different positions of the central ``particle/soliton'' at $z = \SI{500}{km}$ depicted in (a).}
	\label{fig:3sol}
\end{figure}

The initial setting is depicted in Fig.\,\ref{fig:3sol}(a) and consists of 3 unequally spaced solitons with no differential velocity (but with a weak velocity with respect to the trap). By finely tuning the parameters of the particle model (position and amplitude of the solitons) a fairly good overall agreement with the experiment is reached as can be seen in Fig.\,\ref{fig:3sol}(b) (Note that we impose the same amplitude for the 3 solitons for simplicity which is not exactly realized in practice). We stress however the sensitivity of this agreement to the initial positioning of the central particle. Indeed, a small displacement of the latter in the initial state rapidly alters the computed trajectories (Fig.\,\ref{fig:3sol}(c)) up to a very significant qualitative change where the particle model predicts the formation and complex evolution of two-solitons bound-state subsequent to the simultaneous interaction of the 3 components (Fig.\,\ref{fig:3sol}(d) at $\SI{2300}{km}$).  

Such behaviour is not observed nor expected in our work because the integrable features of soliton collision are globally preserved within the range of parameters of the experiments. Specifically, the outcome of collisional processes involving 3 solitons even simultaneously is in agreement with the sum of the effect of the 3 constitutive pairwise interactions \cite{copie_spacetime_2023}.

Finally, we note that several details of the experiments might easily hinder accurate correspondence with the Hamiltonian model including the fact that the pulses are not exactly fundamental solitons, the continuous interaction with radiative content, and non-measurable deformation of the external potential.    


\end{document}